\documentclass{article}
\begin{document}

\begin{title}
\centerline{\large \bf Alternative explanation of the nature of line clumps}
\end{title}

\vspace{3 pt}
\centerline{\sl V.A.Kuz'menko}
\vspace{5 pt}
\centerline{\small \it Troitsk Institute for Innovation and Fusion Research,}
\centerline{\small \it Troitsk, Moscow region, 142190, Russian Federation.}
\vspace{5 pt}
\begin{abstract}

	It is assumed, that the clumps of lines do not connected with states
mixing and IVR, but they are the result of breaking (destruction) of the
process of averaging of momentum of inertia of molecules during the vibration
motion of atoms. Rough estimates of the widths of clumps of lines in
absorption spectra of some acetylenic derivatives were made with this model.
Obtained results are in a satisfactory agreement with the available
experimental data. This idea allows also in principle to explain the origin
of intensive wings of lines, the existence of which was discussed earlier.

\vspace{5 pt}
\end{abstract}

\vspace{12 pt}

	It is known that single lines in the absorption infrared spectrum of
polyatomic molecules can split into a clump (a cluster) of closely spaced
narrow lines [1]. It is supposed that this results from mixing of the states.
Recently the structure of line clumps for various molecules has been studied
rather intensively [2]. On the basis of these data the time of intramolecular
vibrational relaxation (IVR) is calculated.

	The Doppler width of individual lines in a clump is determined only by
the potentialities of the experimental technique in use and it can equal
several MHz. This corresponds to the lifetime of the probed states equal to
several dozens nanoseconds. At the same time, from the analysis of the
structure of line clumps a conclusion is drawn that the IVR duration is, for
example, about 200 ps. No weighty arguments are available which could
eliminate this contradiction. Moreover, except a not very good correlation
between the densities of lines in a clump and the state densities of a
molecule no other reliable data are available to prove that clumps are
formed as a result of mixing of states rather than of some other reasons.
Previously such proofs were not necessary. This explanation was considered
as evident and having no alternative.

	The purpose of the present note  is to suggest such alternative 
explanation of the nature of line clumps which is associated neither with 
mixing of the states nor with IVR. Spectroscopy allows high precision 
(\textless 0.01\% ) determination of the inertia moment of a molecule. But 
this is an averaged inertia moment, because of an inertia moment of molecule 
change during a vibrational motion of atoms in the range $\sim 5 \% $ . Thus, 
we have a reliably established experimental fact: averaging of the inertia 
moment of molecule takes place during the vibration period of their atoms. If 
one supposes that the mechanism of averaging the inertia moment can suffer 
periodic short--time breaking (destruction) by various vibrational modes then
instead of one narrow line in the spectrum we should get  something very 
similar to a clump of lines. The distribution of the lines inside the clump
can be expected to be rather chaotic. And the width of the clump of lines 
is apparently determined by the maximum amplitude of the variation of the 
molecule inertia moment under atom vibrations.

	Clump widths in accordance with the present model can be roughly 
evaluated using the available data on the amplitude of atom vibrations for
various chemical bonds [3]. If one takes into account that in the expression 
for the molecule inertia moment we have the square interatomic distance 
$({\bf R})$ and that the amplitude of atoms vibrations 
$({\scriptstyle\Delta} {\bf R})$ is relatively small the expression for the 
width of clumps $(\bf \Delta )$can be written as:

\begin{equation} 
\bf\Delta=2\bf{B} \Bigl[\biggl(\frac {\bf{R}+
\scriptstyle{\Delta}\textstyle\bf{R}}{\bf{R}}\biggr)^2 -1\Bigr]\approx 
4\bf{B} \frac{\scriptstyle\Delta\textstyle\bf{R}}{\bf{R}}
\end{equation} 

where {\bf B } is the rotational constant of the molecule.

	In Table 1 the results of the appropriate calculations for the width
of line clumps for some acetylenic derivatives are presented. For lighter
molecules (propyne and 1-butyne) the C-H bond parameters were used as a 
model for calculation, whereas for $CF_3 CCH$ and $SF_5 CCH$ molecules the 
parameters of C--F and S--F bonds, respectively, were used. One should keep in 
mind that the data on the vibration amplitude used in Table 1 were taken from 
Ref.[3] and they characterize the mean square vibration amplitude. For the 
width of the line clumps the total vibration amplitude is, apparently, more 
important. Therefore, the calculation results presented in Table 1 can be 
considered $\sim \sqrt2 $ times underestimated.
\begin{center}

{\bf Table 1. Evaluation of the clump widths of the acetylenic C-H stretch  
vibration in some acetylenic derivatives.}

\vspace{8pt}

\begin{tabular}{|l|c|c|c|c|c|}                       \hline
                   &        &      &       &        &            \\
{\bf Molecule} & ${\bf B}^a$ & ${\bf R}^b$ & ${\scriptstyle\Delta} \bf R$ &
${\bf \Delta}^c_{calc}$ & ${\bf \Delta}^d_{exp} $                \\
&$(cm^{-1}) $ &  (\.{A})  &(\.{A})  &$(cm^{-1})$& $(cm^{-1})$ \\ \hline
                   &        &      &       &        &             \\
  $CH_3 CCH $ 	   & 0.284  & 1.08 & 0.078 & 0.082  & $\sim0.10$  \\ 
                   &        &      &       &        &             \\ \hline
                   &        &      &       &        &             \\
  $CH_3 CH_2 CCH $ & 0.144  & 1.08 & 0.078 & 0.042  & $\sim0.06$  \\
                   &        &      &       &        &             \\ \hline
                   &        &      &       &        &             \\ 
  $CF_3 CCH $ 	   & 0.0957 & 1.33 & 0.045 & 0.013  & 0.02--0.08  \\ 
                   &        &      &       &        &             \\ \hline
                   &        &      &       &        &             \\
  $SF_5 CCH $ 	   & 0.0579 & 1.58 & 0.042 & 0.0062 & 0.002--0.008\\
                   &        &      &       &        &             \\  \hline
 
\end{tabular}
\end{center}
\vspace{5pt}

a-- the values of the rotational constant {\bf B} were taken from Ref.[4-7]. 

b-- the values of interatomic distances were taken from Ref.[8].

c-- results of calculations by the formula (1)

d-- the values of the clump widths were taken  from Ref.[5-7,9].

	Thus, it is seen from Table 1 that the results of even a rather rough
evaluation  of the clump widths within the framework of the present model
agree satisfactorily with the experimental results. 

	Along with clumps of lines in the spectra of polyatomic molecules
purely Lorentzian line contours [10] of comparable width are also observed. 
The authors of the works usually do not see any principal differences between 
them. In our opinion, the purely Lorentzian contour actually characterizes 
the IVR process. Therefore, it cannot have a fine inner structure and, in 
fact, has nothing in common with line clumps. It is interesting that in 
Ref.[6] one can observe a gradual transition from a clumps of lines to the 
purely Lorentzian contour: this occurs in this case at the growing number of 
the rotational transition. 

	Investigation of the structure of line clumps does not yield now rich 
information. On the one hand, now only rather trivial information can be 
obtained from the clump widths. But on the other hand, it is difficult to 
understand as yet how to interpret definitely the data on the density of 
lines in a clump. In our opinion, it is much more interesting to study the 
wings of absorption lines for polyatomic molecules. Quite reliable 
experimental proofs are available to show that narrow absorption lines of 
polyatomic molecules have wide and intensive Lorentzian wings of unknown 
nature [11]. Namely the breaking (destruction) of the averaging process of 
the inertia moment of the molecules can, in principle, be responsible for 
both: the appearance of clumps and the intensive line wings. The very moment 
of breaking (destruction) can be  considered as some short-life  state of the 
molecule, which should  be corresponded with some wide component in the line 
spectrum. 

A sufficiently reliable measurement of the intensity and of the width of the 
line wings can easily be performed in a molecular beam by the technique 
available. Molecular beam with a cryogenic bolometer is ideally suitable for 
the study of line wings [12].  Narrow and powerful laser radiation interacts 
mainly with the line wings. Low rotational temperature in the beam  results 
in that the width of  the absorption band is comparable with or less 
than that of the line wings. Therefore, for precise determination of the width
of the line wings  it is  not necessary to investigate very far wings of the 
absorption bands.  The sensitivity of the apparatus can easily be calibrated
using large polyatomic molecules like as hexafluoroacetone. Upon calibration,
such an instrument can be used for measuring the intensity and the width of 
the line wings for most  polyatomic  molecules. This information is  important
for the solution of such spectroscopic problems as infrared multiphoton 
excitation of molecules, infrared laser dissociation of dimers and clusters, 
the nature of far wings in absorption bands of molecules in the gaseous phase,
the line widths for molecule absorption in liquids, etc. It is quite possible,
that the Earth's climate is depended on such intense line wings [13,14].

\end{document}